\documentstyle[twocolumn,psfig,aps]{revtex}
\begin{document}
\draft
\twocolumn[\hsize\textwidth\columnwidth\hsize\csname @twocolumnfalse\endcsname
%
%
%

\title{Hole-Pairs in a Spin Liquid:
Influence of Electrostatic Hole-Hole Repulsion}


\author{Claudio Gazza$^1$, George B. Martins$^1$, Jos\'e Riera$^{2,*}$, and Elbio Dagotto$^1$}

\address{$^1$ National High Magnetic Field Lab and Department of Physics,
Florida State University, Tallahassee, FL 32306}
\address{$^2$ Laboratoire de Physique Quantique \& Unit\'e mixte
  10  de Recherche CNRS 5626, Universit\'e Paul Sabatier,
  11  31062 Toulouse, France}


\date{Submitted for publication on March 1998}
\maketitle

\begin{abstract}
The stability of hole bound states in the $t-J$ model including
short-range Coulomb interactions is analyzed using computational
techniques on ladders with up to $2 \times 30$ 
sites. For a nearest-neighbors (NN)
hole-hole repulsion, the two-holes bound state is surprisingly robust and
breaks only when the repulsion is several times the exchange $J$.
At $\sim 10\%$ hole doping the pairs break only for a NN-repulsion as
large as $V \sim 4J$. Pair-pair correlations remain robust in the regime
of hole binding. The results support electronic hole-pairing mechanisms on
ladders based on holes moving in spin-liquid backgrounds. 
Implications in two dimensions are also
presented. The need for better estimations of the range and strength of
the Coulomb interaction in copper-oxides is remarked.

\end{abstract}
\vskip2pc]
\narrowtext

%
%

The discovery of superconductivity in 
$Sr_{14-x} Ca_x$ $Cu_{24}O_{41}$~\cite{uehara} under high pressure
has triggered a considerable effort to
understand the physics of copper-oxide ladder materials. Early theoretical
studies predicted the existence of superconductivity based on a
purely electronic mechanism~\cite{science}.
The rationale for the hole-hole  attraction
is that each individual hole 
distorts the spin-liquid ground state of the undoped
ladder, increasing locally the energy density.
The damage to the spin
arrangement is, thus, minimized if holes share a common 
distortion. 
Several  calculations, including computational
studies for the $t-J$ model~\cite{science,review,tsunetsugu,didier},  
have confirmed 
that  an effective attraction between holes exists in 
environments with short-range AF correlations. 
The fact that the
 hole-pairs transform as $d_{x^2 - y^2}$ under rotations, i.e.
the same channel in which the real doped 2D cuprates
superconduct, provides additional support to the
 electronic mechanisms described above.

However, studies of hole pairing in spin liquids
 are usually performed without the introduction of
an intersite hole-hole Coulomb repulsion.
The accuracy of such an approximation has been much
debated, but there are few robust calculations supporting the various points
of view. The neglect of
electrostatic interactions is particularly important for 
 values of $J/t$ such as 0.3-0.4, presumed to be realistic, since in
this regime the pair size is small, roughly $2a$ (where $a$ is the
lattice spacing)~\cite{tsunetsugu,didier,white1}. 
Nevertheless, note that in the real materials at least
the 5 d-orbitals
of $Cu$, as well as the 3 p-orbitals of $O$, should be taken 
into account. In this extended model,
 polarization effects caused by electron-hole excitations 
can effectively reduce the strength of the 
electrostatic repulsion. Actually, 
$O^{2-}$ has a large polarizability 
that influences on the effective Coulomb
interactions of a variety of ionic insulators~\cite{ishi}.
This effect has been analyzed for the
on-site repulsions $U_{d}$ and 
$U_{p}$~\cite{sawa}. Although the bare couplings are
substantially reduced by polarization,
their strength
remains the dominant scale and models
with no doubly occupied sites capture properly this effect.
However, it is unclear if the additional neglect of
the hole-hole repulsion at distance $a$
remains a good approximation.

Unfortunately, the explicit
calculation of the strength of screened
Coulombic interactions is difficult. 
Using results for 2D copper-oxides 
as guidance, a constrained-density-functional approach where the LDA bands are
identified with a mean-field solution of the Hubbard model
reports a  repulsion $U_{pd} \sim 1.2 eV$ between a pair of holes
at the $Cu$ and $O$ ions~\cite{hyb}. Since the bare value is
$7.8 eV$, a reduction of a factor 6.5 is effectively achieved by
polarization effects. If the trend continues, then
the repulsion $V$ at distance $a$ 
will be $\sim 0.6 eV \sim 4J$ (with $J\sim 0.15 eV$~\cite{hayden}).
However, other estimations comparing Auger
spectroscopy results 
with
${(Cu_4 O_5)^{6-}}$ cluster calculations using a multiband Hubbard
model report 
$U_{pd} < 1eV$~\cite{sawa}. In addition, for two holes in neighboring oxygens
$U_{pp} \sim 0$ was observed~\cite{sawa},
 suggesting that the hole-hole correlations
decay rapidly with distance. For this reason the above estimation
$0.6 eV$ should be
considered as an upper bound for $V$.
%
%
%
Other studies based on  two
neighboring Zhang-Rice singlets report $V \sim 0.2 eV$~\cite{other}.
Finally, approaches where
the  dielectric constant
$\epsilon \sim 30$~\cite{chen} is 
used even at short distances
provide $V \sim 0.13eV$. Then, current estimations locate
$V \epsilon [J,4J]$ as the realistic range for the hole-hole
repulsion in a one band model.
Such strong repulsion is dangerous for real-space AF theories.
Coulomb interactions are also a central
ingredient of the ``striped'' scenarios~\cite{emery}.

The purpose of this paper is to discuss a computationally intensive
calculation of the effect of intersite hole-hole repulsion on the hole bound
states of the $t-J$ model on ladders. Since the
size of the hole pairs on planes and ladders is 
comparable for the same 
$J/t$, our results have consequences also for studies in 2D.
The calculation is performed
with the DMRG technique~\cite{white}, 
as well as using a
diagonalization technique in a reduced Hilbert space~\cite{oldtrunca}
(the Optimized Reduced-Basis Approximation or ORBA)~\cite{change}. The
Hamiltonian employed here is
the $t-J$ model at a realistic coupling $J/t=0.4$,
 supplemented by a hole-hole repulsion
$V_{hh} = \sum_{\bf ij} (V/|{\bf i} - {\bf j}|)  n_{\bf i} n_{\bf j}$, where
$n_{\bf i}$ is the hole number operator at site ${\bf i}$. The
range $R$ of the repulsion has been restricted to  1 and $\sqrt{2}$
lattice spacings,
since the speed of 
convergence of the numerical techniques (both variational)
decreases as $R$ grows. Earlier results have been
obtained in the case of 
static holes~\cite{barnes} or using small clusters~\cite{previous}.

Let us discuss first the case of two holes.
Fig.1 contains the hole-hole correlation $C({\bf j}) 
= \langle n_{\bf 0} n _{\bf j} \rangle$
obtained on $2 \times N$ clusters with two holes and open boundary conditions (OBC)
using DMRG ($m=300$ states),
for the case where $V_{hh}$ acts only at a
distance of one lattice spacing (i.e. $R=1$)~\cite{comm3}.
${\bf 0}$ is a site at the center of the cluster, and the figure shows the
hole-hole correlation along the leg opposite to where ${\bf 0}$ is located
(results for the other leg are similar). $C({\bf j})$ 
is related with the probability of finding one hole at ${\bf j}$
when there is one at ${\bf 0}$.
\vspace{-0.6cm}
\begin{figure}
\psfig{figure=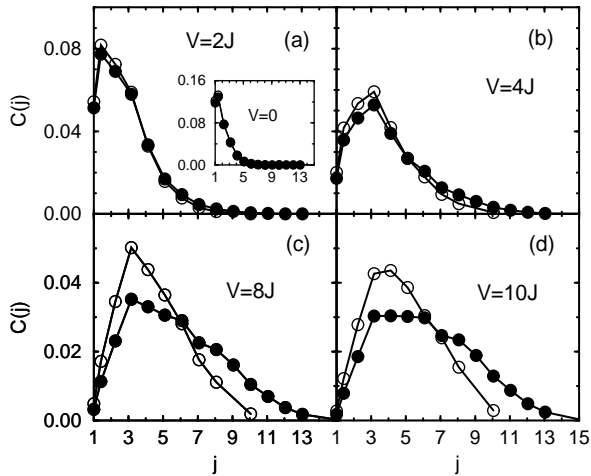,height=6.8cm,width=8.0cm,angle=-90}
\vspace{0.4cm}
\caption{
Hole-hole correlation $C({\bf j})$ (see text) for
$2 \times 20$ (open circles) and $2 \times 30$ (full circles)
clusters with two holes,
using DMRG with OBC and $J/t=0.4$.
Results for several couplings $V/J$ are
shown. The Coulomb repulsion has range $R=1$.
Here ${\bf j}$ is the distance between a site
at the center defined as origin, and sites belonging to the opposite
leg.
}
\label{fig1}
\end{figure}
Figs.1a-b correspond to $V=0$, $2J$ and $4J$.
Here the results change only slightly as the lattice
grows, and the two holes remain  close to each
other indicating the existence of a bound state. Apparently  a 
NN-repulsion already larger than $J$ is $not$
enough to destroy the bound state, although it weakens it.
On the other hand, Figs.1c-d show 
similar results but now for $V=8J$ and $10J$ where a
substantial change in the hole distribution is observed
as the cluster grows. The spreading of the hole over the entire
lattice suggests either the breaking of the pair
or a weak bound state.
In the large-$V$ regime one of the holes acts
as a sharp ``wall'' to the other, which spreads its wave
function  in an
effective square-well potential.
Fig.2a shows an average hole-hole distance defined as $ \langle d
\rangle
= \sum_{\bf j \neq 0} d_{\bf j} \langle n_{\bf 0} n_{\bf j} \rangle
 / \sum_{\bf j \neq 0} \langle n_{\bf 0} n_{\bf j} \rangle $,
where $d_{\bf j}$ is the ${\bf 0}$-${\bf j}$ distance. The convergence
of the results for $V \leq 4J$  as the size grows
is clear. However, for $V \geq 6J$,
there is no obvious convergence.
The curvature change of the $2 \times 30$ results
at $V \sim 5J$ also favors the interpretation that the bound state
exists up to that coupling,
since in the bulk $\langle d
\rangle$ vs $V/J$ will grow with positive curvature,
diverging at the pair-breaking critical coupling.
\vspace{-0.95cm}
\begin{figure}
\psfig{figure=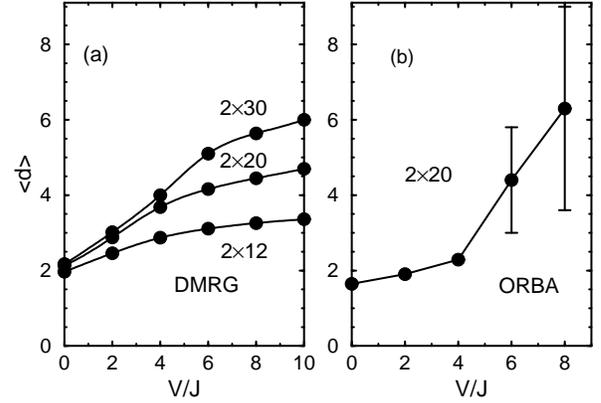,height=6.6cm,width=8.0cm,angle=-90}
\caption{
(a) Mean distance between holes
$\langle d \rangle$, defined in the text, vs
$V/J$ obtained with the DMRG method at $J/t=0.4$.
Results for three lattice sizes are shown.
The range of the Coulomb repulsion is $R=1$; (b)
Same as (a) but now using the ORBA technique with
$\sim 10^6$ states, and a $2 \times 20$ cluster with periodic boundary
conditions.
}
\label{fig2}
\end{figure}
To further support these conclusions, the ORBA technique has been
implemented on a $2 \times 20$ cluster with periodic boundary
conditions (PBC). The basis selected to optimize the  convergence as
the Hilbert space grows is the rung-basis, which contains 9 states for
the $t-J$ model~\cite{change}. 
A convergence in the energy with up to 3 significant
figures is achieved in this cluster by using $\sim 10^6$ states
in the subspace of zero momentum. 
While this is 
not as precise as the DMRG method for this
model,
the accuracy is enough for our
purposes and ORBA has
the advantage that it is performed with PBC.
Fig.2b shows $\langle d
\rangle$ vs $V/J$. Here ${\bf 0}$ is any site of the lattice since there is
translational invariance. For $V/J \leq 4$ it was observed that the
 hole-hole correlations change by a very small percentage when
 the space grows from $10^5$ to $10^6$ states, using as a starting
configuration two holes in the same rung.  In this regime the
hole-pair bound state is tight.
However, for $V/J \geq 6$ the changes in the correlations
 as the number of states grows are large.
Actually 
the hole-hole distances reported in Fig.2b for large $V/J$ are not
stabilized 
(thus they carry large error bars).
Nevertheless, as $V/J$ increases
the qualitative behavior of 
$\langle d \rangle$  is in 
 agreement with the DMRG data.



Let us now analyze
what occurs when the range of the repulsion grows. 
Fig.3 contains DMRG results for a repulsion 
of strength $V$ and $V/{\sqrt{2}}$ when holes are at distances 1
and $\sqrt{2}$ lattice spacings, respectively, 
and zero otherwise. Although qualitatively
similar to Fig.1, now the region where the bound state is stable
is restricted to $V/J \leq 2$, while results for $V/J \geq 4$ suggest
the absence of a bound state due to the spreading of the second hole
when one is  at the center of the ladder.
Fig.4a shows $\langle d \rangle$ for the same interaction  used in
Fig.3. In agreement with the previous discussion, the critical region
is $V/J \sim 3$. For smaller (larger) couplings $\langle d
\rangle$  converges (diverges) as the cluster size grows. 
ORBA results 
also suggest a
critical coupling $V/J \sim 3$.
\vspace{-0.6cm}
\begin{figure}
\psfig{figure=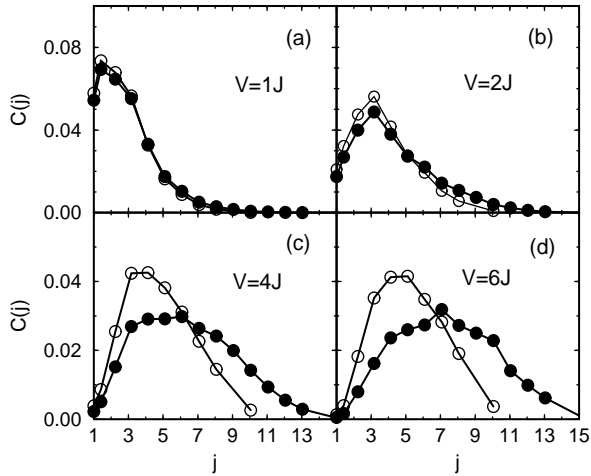,height=6.8cm,width=8.0cm,angle=-90}
\vspace{0.4cm}
\caption{
Same as Fig.1 but with a longer ranged repulsion of strength $V$
($V/\sqrt{2}$) at distance of one ($\sqrt{2}$) lattice spacings.
}
\label{fig3}
\end{figure}

To complement the analysis using hole-hole correlations,
Fig.4b contains the binding energy defined as
 $\Delta_B = E(2) + E(0) - 2 E(1)$,
where $E(n)$ is the lowest energy in the subspace of $n$ holes. A
negative
$\Delta_B$ implies that a bound state exists. 
At $J/t=0.4$ and $V=0$
there is pairing of holes on 2-leg ladders~\cite{science,tsunetsugu}.
The figure shows that for
a NN-repulsion, the binding effect continues even
 up to large
values of $V/J$. Actually the region $6 \leq V/J \leq 10$
corresponds to a weak bound state.
The same figure shows results for range-$\sqrt{2}$ repulsion.
In
this case the critical coupling $V/J |_c$ at which the bound state is lost is
between 3 and 4. Both results are in agreement with the estimations based
on Figs.3-4a.

The previous analysis indicate that
the stability of the two-hole bound states in $t-J$-like models 
depends  on the value and range of the electrostatic Coulomb
interaction between holes.
When the Coulombic term is restricted to a realistic range $R=1$, 
a repulsion as large as $0.6
eV \sim 4J$ weakens but does not destroy the pair, implying that
the effective
range of attraction caused by spin polarization
is larger than one lattice spacing. Retardation effects (fully
considered in the present calculation) due to the
different energy scales 
of spin and charge excitations ($J$ vs $t$) likely contributes
to the strong stability of the bound states~\cite{previous}.
Note also that using, e.g., $V/J=4$, the pair size is $\sim 4-5a$ (Fig.2a).
This result is close to estimations of the Cooper pair size in the
2D cuprates which report a coherence length $\xi_c/a \sim 4$ and $\sim 3$,
for optimally doped $La-214$ and $YBCO$, respectively~\cite{cyrot,muon}
(using $a=3.8 \AA$). To the extend that $\xi_c/a$ are similar on
planes and ladders, apparently a realistic NN hole-hole repulsion can 
actually improve
quantitatively the predictions of the $t-J$ model, without destroying
the pairs.

\vspace{-1.0cm}
\begin{figure}
\psfig{figure=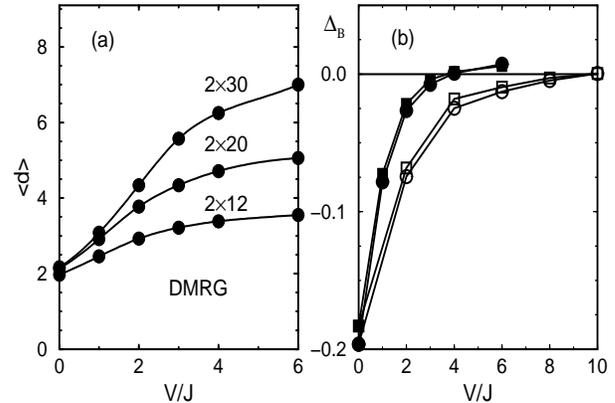,height=6.6cm,width=8.0cm,angle=-90}
\caption{
(a) Same as Fig.2a
but with a longer ranged repulsion of strength $V$
($V/\sqrt{2}$) at
distance of one ($\sqrt{2}$) lattice spacings;
(b) DMRG binding energy of two holes obtained 
for $J/t=0.4$. Open circles and squares
correspond to
$2 \times 20$ and $2 \times 30$ clusters, respectively, both with a
Coulomb repulsion of range $R=1$. Full circles and
squares also correspond to
$2 \times 20$ and $2 \times 30$ clusters, respectively, but now with
a Coulomb repulsion of range $R=\sqrt{2}$.
}
\label{fig4}
\end{figure}



The results discussed thus far have been restricted to the two holes
sector. Since the pairs are very tight it is reasonable to assume that
even with a finite (small) hole density the bound states will remain stable
upon the addition of a NN-repulsion $V$. This assumption
can be checked explicitly
using a $2 \times 20$ cluster (shown to be already close to the bulk
limit, see Fig.4b). In Fig.5a the binding energy  defined as
$\Delta_B = E(n) + E(n-2) - 2 E(n-1)$ vs the hole concentration
$x = n/40$ is shown for three values of $V$. As expected, the results are
qualitatively similar to those of the two holes problem. While for $V=0$
the pairs remain stable even up to 20\% doping and beyond, using a
repulsion as large as  $V=2J$
the stability persists roughly up to  $x \sim 0.17$, 
and for $V=4J$ up to
$x \sim 0.10$. Finding a critical coupling $V/J|_c$ decreasing
with doping is reasonable since the stability of the pairs at finite
$x$ depends not only on the hole-hole attraction, but also on the mean
distance between holes that introduces limitations on the pair size.
As example consider $V/J=2$. From Fig.2a the
pair size is $\langle d \rangle \sim 3$. The density at which the hole mean-distance is
also $\sim 3$ corresponds to $x \sim 0.17$,
in good agreement with the critical density at this $V$ (Fig.5a).
Then, from the two-hole problem information about finite $x$ can be
obtained.  

Superconducting pair-pair correlations $C(r)$ have also been measured
here.  $C(r)$ is proportional to
$ \sum_m \langle \Delta(m) \Delta^\dagger(m+r) \rangle$, where $m,r$
are rung indices, and $\Delta(m)$ destroys a pair of electrons in a spin singlet
 at rung-$m$. An average over the whole cluster is used to calculate $C(r)$.
At $V=0$ previous results were reproduced as a
test~\cite{hayward}. The new results are shown in Fig.5b using hole-density $x=0.1$
as example. The data is presented at three characteristic distances
($r=3,6,10$)~\cite{super} vs $V/J$. The normalization using $C(0)$ is
important since the signal for superconducting correlations has an
overall penalization due to the reduction of the probability of finding
two holes in the same rung as $V/J$ grows. This reduction only indicates that
pairs become larger as $V/J$ increases.
Fig.5b shows that the strong $V=0$
superconducting correlations remain robust
as $V/J$ grows. Even for $r$ as large as
10 the correlations are not negligible in the scale of the plot in the
region $0 \leq V/J \leq 4$, which coincides with the region of hole
binding (Fig.5a). Similar conclusions were
reached at $x=0.2$ (not shown). Then, our results are compatible with a
picture where hole-binding and strong superconducting correlations
coexist. Note, however, that as $x\rightarrow 0$ the pair correlations
must decrease due to the reduction in the density of pairs even if 
the binding energy is robust.

\vspace{-1.3cm}
\begin{figure}
\psfig{figure=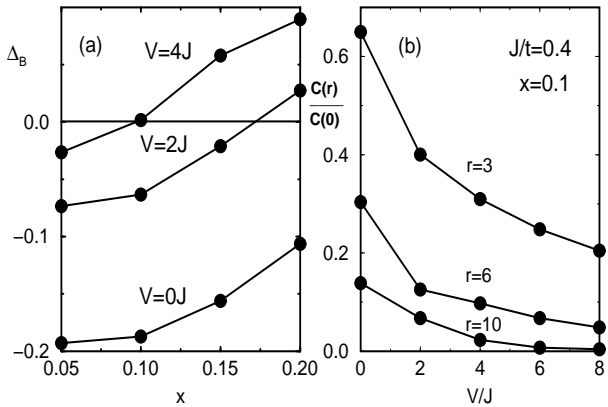,height=6.6cm,width=8.0cm,angle=-90}
\caption{
(a) Binding energy of hole pairs obtained working at a finite density of holes on
a $2 \times 20$ cluster. Results at $J/t=0.4$ and several $V$'s are
shown;
(b) Pairing correlations $C(r)/C(0)$ (see definition in text) vs $V/J$
at three characteristic distances $r$ along the legs. $J/t=0.4$ and the
hole density is $x=0.1$. The cluster size is $2 \times 30$.
}
\label{fig5}
\end{figure}


Summarizing, it has been shown that (1) the two-holes ladder
bound state 
is more stable than expected
upon the introduction of a NN hole repulsion;
and that (2) this conclusion persists in the presence of a finite
hole density. Hole pairs and robust
superconducting correlations were found in the same region of parameter space.
These results provide support to ladder theories that predict
hole-pairing based on electronic mechanisms that describe holes as
immersed in spin-liquid backgrounds~\cite{science}. It 
also reinforces striped scenarios for
cuprates where pairing is produced by carriers moving from the
fluctuating stripes
to the ladders between them~\cite{emery}. 
Regarding 2D systems our results show that the effect of
a Coulomb interaction is a subtle quantitative problem. Actually the
binding energy of two holes on a 4-leg ladder with $V=0$ and $J/t=0.4$ 
has also been
estimated here. The result is $|\Delta^{4L}_B| \sim 0.14t$ (using a $4 \times
12$ cluster, $m \sim 400$ states, and truncation error $\sim 10^{-5}$)
which is similar to previous
results in 2D (Fig.24 of Ref.~\cite{review}).
Then, the influence of the NN Coulomb interaction on planes will
be stronger than on ladders.
In this respect it is imperative to obtain
either experimental or theoretical information about the range and 
strength of the Coulomb interaction in effective one-band models for 
cuprates, specially in 2D.



Support of the NSF grant DMR-9520776,  CONICET (Argentina),
CNPq-Brazil, and the NHMFL
In-House Research Program (DMR-9527035) is acknowledged. 
The authors thank V. Emery, S.
Kivelson, S. Maekawa, and G.
Sawatzky for useful comments.

\medskip

\vfil


\vspace{-0.8cm}
\begin{references}

\vspace{-1.6cm}

\bibitem[*]{} Permanent address: Inst. de 
  F\'{\i}sica Rosario, 
  Univ. Nac. de Rosario, Av. Pellegrini 250, 2000-Rosario,
  Argentina.

\bibitem{uehara} M. Uehara, et al.,
J. Phys. Soc. Jpn. {\bf 65}, 2764 (1996).

\bibitem{science} 
E. Dagotto and T. M. Rice, 
Science {\bf 271}, 618 (1996).




\bibitem{review} E. Dagotto, Rev. Mod. Phys. {\bf 66}, 763
(1994).

\bibitem{tsunetsugu} H. Tsunetsugu et al.,
Phys. Rev. {\bf B 49}, 16078 (1994).

\bibitem{didier} D. Poilblanc et al., 
Phys. Rev. {\bf
B 49}, 12318 (1994). See also D. Poilblanc,
Phys. Rev. {\bf B 49}, 1477 (1994).

\bibitem{white1} S. White and D. Scalapino, Phys. Rev. {\bf B 55},
6504 (1997), and references therein.



\bibitem{ishi}
 S. Ishihara, et al., 
Phys. Rev. {\bf B 49}, 16123 (1994).

\bibitem{sawa} L. H. Tjeng, H. Eskes, and G. A. Sawatzky, in {\it Strong
Correlation in Superconductivity}, Eds. H. Fukuyama, S. Maekawa, and A.
P. Malozemoff, Springer series in Solid-State Sciences, Vol. 89 (1989).

\bibitem{hyb} 
M. S. Hybertsen, et al., Phys.
Rev. {\bf B 39}, 9028 (1989).

\bibitem{hayden} S. M. Hayden et al., Phys. Rev. Lett. {\bf 67}, 3622 (1991).


\bibitem{other} F. Barriquand and G. A. Sawatzky, Phys. Rev. {\bf B 50},
16649 (1994).

\bibitem{chen} C. Y. Chen et al., Phys. Rev. Lett. {\bf 63}, 2307
(1989). 

\bibitem{emery}
 V. J. Emery, S. A. Kivelson, and O. Zachar, Phys. Rev.
{\bf B 56}, 6120 (1997). These authors also
estimated the bare Coulomb repulsion at $R=1$
as $0.5 eV$.








\bibitem{white} S. R. White, Phys. Rev. Lett. {\bf 69}, 2863 (1992).

\bibitem{oldtrunca} J. Riera and E. Dagotto, Phys. Rev. {\bf B 47}, 15346
(1993); {\it ibid} {\bf B 48}, 9515 (1993);
and references therein.

\bibitem{change} E.~Dagotto, et al., to appear in Phys. Rev. {\bf B}.




\bibitem{barnes} 
T. Barnes and M. Kovarik, Phys.
Rev. {\bf B 42}, 6159 (1990). 


\bibitem{previous} 
J. Riera and E.
Dagotto, Phys. Rev. {\bf B 57}, 8609 (1998).


\bibitem{comm3} 
The
 truncation error $(1 - P(m))$ was $\sim 10^{-5}$ or better. 



\bibitem{cyrot} M. Cyrot and D. Pavuna, {\it Introduction to
Superconductivity and High-Tc Materials}, World Scientific (1992).

\bibitem{muon} 
Recent estimations by
B. Nachumi et al., Phys. Rev. Lett. {\bf 77}, 5421 (1996) arrive to
$\xi_c = 18.3 \AA$ and $22.7 \AA$ for 1\%
Zn-doped underdoped $YBCO$ and $La-214$, respectively.





\bibitem{hayward}  
C.~Hayward, et al., Phys. Rev. Lett. {\bf 75}, 926 (1995).
%

\bibitem{super}Results at distances larger than 10 carry large errors.





















\end{references}
\end{document}